\documentclass[12pt]{article}
\usepackage{amssymb}

\begin{document}

\begin{titlepage}
\begin{center}

\vspace*{10mm}

{\LARGE\bf
Radiative Symmetry Breaking on D-branes
at Non-supersymmetric Singularities}

\vspace*{20mm}

{\large
Noriaki Kitazawa
}
\vspace{6mm}

{\it
Department of Physics, Tokyo Metropolitan University,\\
Hachioji, Tokyo 192-0397, Japan\\
e-mail: kitazawa@phys.metro-u.ac.jp
}

\vspace*{15mm}

\begin{abstract}
The possibility of radiative gauge symmetry breaking
 on D3-branes at non-supersymmetric orbifold singularities
 is examined.
As an example,
 a simple model of D3-branes
 at non-supersymmetric ${\bf C}^3/{\bf Z_6}$ singularity
 with some D7-branes for the cancellations of R-R tadpoles
 in twisted sectors is analyzed in detail.
We find that
 there are no tachyon modes in twisted sectors,
 and NS-NS tadpoles in twisted sectors are canceled out,
 though uncanceled tadpoles and tachyon modes
 exist in untwisted sectors.
This means that
 this singularity background
 is a stable solution of string theory at tree level,
 though some specific compactification of six-dimensional space
 should be considered for a consistent untwisted sector.
On D3-brane
 three massless ``Higgs doublet fields'' and
 three family ``up-type quarks'' are realized at tree level.
Other fermion fields, ``down-type quarks'' and ``leptons'',
 can be realized as massless modes of the open strings
 stretching between D3-branes and D7-branes.
The Higgs doublet fields
 have Yukawa couplings with up-type quarks,
 and they also have self-couplings
 which give a scalar potential without flat directions.
Since there is no supersymmetry,
 the radiative corrections may naturally develop
 negative Higgs mass squared and ``electroweak symmetry breaking''.
We explicitly calculate
 the open string one-loop correction to the Higgs mass squared
 from twisted sectors,
 and find that the negative value is indeed realized
 in this specific model.
\end{abstract}

\end{center}
\end{titlepage}

\section{Introduction}
\label{intro}

Although many phenomenological models
 using D-branes in string theory have been constructed
 (see for review
  Refs.\cite{Blumenhagen:2005mu,Kokorelis:2004tb}
  and references therein),
 few attention has been paid on the dynamics of
 gauge symmetry breaking: the electroweak symmetry breaking,
 grand unified gauge symmetry breaking, and so on.
In Ref.\cite{Antoniadis:2000tq}
 the possibility of radiative electroweak symmetry breaking
 has been examined by the explicit string calculation
 in a specific ``brane supersymmetry breaking'' model
 \cite{Antoniadis:1999xk}.
In the model
 the Wilson line in a compact direction of the radius $R$
 is identified as a scalar field,
 and there is no potential for the scalar field at tree level. 
Because of no supersymmetry,
 the potential of the scalar field can emerge at one-loop level.
It has been shown by an explicit calculation in string theory that
 the scalar field can have finite vacuum expectation value
 roughly of the order of $1/R$.
Though
 the magnitude of the negative mass squared of the scalar field
 depends on both the string scale and $R$ rather a complicated way,
 it is essentially determined by the scale $1/R$
 and one-loop suppression factor.
It has been pointed out,
 through some phenomenological discussions
 on the application to the electroweak symmetry breaking, that
 the string scale can be from TeV scale to the intermediate scale
 depending on the sizes of compact directions.
In spite of this interesting possibility
 of radiative electroweak symmetry breaking, however,
 not so much subsequent efforts
 towards constructing realistic models have been made
 in the framework of string theory.

One of the difficult problems of this direction
 is that NS-NS tadpoles are not canceled out
 in general in non-supersymmetric models
 \cite{Fischler:1986ci,Fischler:1986tb,Das:1986dy}.
In some D-brane configuration
 R-R tadpoles have to be canceled out
 for consistency reasons (anomaly cancellations).
In supersymmetric configurations
 NS-NS tadpoles are automatically canceled out,
 if the R-R tadpoles are canceled out,
 and the assuming backgrounds (background geometry and flux)
 are solutions of string theory.
This is not the case in non-supersymmetric configurations in general
 and some modifications of the backgrounds are required.
Some one-loop calculations may give divergent results
 due to NS-NS tadpoles without the modification of the background.
The systematic procedure of the modification
 has not yet been established
 (see Ref.\cite{Dudas:2004nd} for recent proposal).

In case that the string scale is in TeV region,
 massive string states may give sizable effects at low energies.
In the low energy effective field theory,
 such effects can be described as some contact interactions
 by the exchanges of massive string states,
 and there is the lower bound in the string scale
 to a few TeV\cite{Antoniadis:2000jv}.
Since
 the size of the six-dimensional compact space
 should be taken appropriately large
 to explain the weakness of the gravitational interaction
 in case of TeV string scale,
 there is a possibility that
 the Kaluza-Klein modes of some fields in the standard model
 are light and give sizable effects at low energies
 \cite{Antoniadis:1990ew,Antoniadis:1993jp}.
Since
 the scenario of TeV string scale predicts these effects
 which can be accessible in future experiments
 (see Ref.\cite{Antoniadis:2005aq} for review),
 it must be very important
 to theoretically examine further the possibility,
 the electroweak symmetry breaking in this paper,
 in the framework of well-defined perturbative string theory.

The outline of this paper is as follows.
In section \ref{model}
 the system of D3-branes
 on non-supersymmetric ${\bf C}^3/{\bf Z}_6$ orbifold singularity
 is discussed in detail,
 and we construct a model in which
 all the R-R and NS-NS tadpoles in twisted sectors are canceled out
 and no tachyons exist in twisted sectors.
This means that
 this non-supersymmetric ${\bf C}^3/{\bf Z}_6$ orbifold singularity
 is a stable solution of string theory.
Three ``Higgs doublet fields'' and ``up-type quark fields''
 are realized on the D3-brane as massless modes of the open string
 at tree level,
 and Higgs doublet fields have tree-level potential
 which has no flat direction.
It is necessary for ``electroweak symmetry breaking''
 that the one-loop radiative correction gives
 Higgs doublet field negative mass squared.
The mass squared of Higgs fields can be obtained
 by calculating one-loop two point function
 in string world-sheet theory.
In section \ref{technique}
 some techniques to calculate two point function
 in string world-sheet theory is reviewed.
In section \ref{results}
 the mass squared of a Higgs doublet field is calculated.
Only the contribution from the twisted sectors are examined,
 since the contribution from the untwisted sector
 is model (compactification) dependent.
We find that
 the mass squared is negative,
 and suggest that
 that the radiative ``electroweak symmetry breaking''
 is possible in this system.
Some additional comments are given
 at the end of this section.

\section{D-branes
 at non-supersymmetric ${\bf C}^3/{\bf Z}_6$ singularity}
\label{model}

We consider
 D3-branes at one ${\bf C}^3/{\bf Z}_6$ orbifold singularity
 which can be considered as one of the singularities
 in some six-dimensional compact space.
Since we examine the properties
 which are determined
 only by the local structure of D-branes at the singularity,
 we consider ${\bf C}^3/{\bf Z}_6$ singularity rather than
 one of the singularities in some concrete compact space,
 say $T^6/{\bf Z}_6$ orbifold, for example.
This is the approach
 which is proposed in Ref.\cite{Aldazabal:2000sa}.

The original massless spectrum on $n$ D3-branes
 is a U$(n)$ gauge multiplet
 of the four-dimensional ${\cal N}=4$ supersymmetry,
 whose components are
 a gauge field $A^{a\mu}$ ($\mu=0,1,2,3$),
 six real scalar fields $X^{a\kappa}$ ($\kappa=1,2,\cdots,6$)
 and
 four Weyl fermion fields $\lambda^a_\alpha$ ($\alpha=1,2,3,4$),
 where $a$ is the index of adjoint representation of U$(n)$.
In the language of four-dimensional ${\cal N}=1$ supersymmetry,
 they consist one ${\cal N}=1$ gauge multiplet
 and three ${\cal N}=1$ chiral multiplets in adjoint representation.
The six real adjoint fields
 can be understood as brane position moduli fields.
Under SU$(4)_{\rm R}$ global symmetry,
 the six real scalar fields and four Weyl fermion fields belong to
 $\bf 6$ and $\bf 4$ representations, respectively.
The transformation on six real scalars
 can be understood as the rotational transformation
 in the transverse six dimensional space.
These fields
 correspond to the states in string world-sheet theory
 in the following way.
\begin{eqnarray}
 A^{a\mu} (T^a)_{{\bar i}j}
 &\sim&
 \psi_{-1/2}^\mu | {\bar i} j \rangle_{\rm NS},
\\
 X^{a\kappa} (T^a)_{{\bar i}j}
 &\sim&
 \psi_{-1/2}^{3+\kappa} | {\bar i} j \rangle_{\rm NS},
\\
 \lambda^a_\alpha (T^a)_{{\bar i}j}
 &\sim&
 | s_1, s_2, s_3; {\bar i} j \rangle_{\rm R}
 \quad
 \mbox{with}
 \quad
 \prod_{r=1}^3 s_r = -1/8,
\end{eqnarray}
 where
 NS and R denote Neveu-Schwarz sector and Ramond sector,
 respectively,
 $\psi^M_{-1/2}$ with $M=0,1,\cdots,9$
 are creation operators of world-sheet fermions,
 and ${\bar i}, j$ are Chan-Paton indices.
Three quantities, $s_r = \pm 1/2$ with $r=1,2,3$ in Ramond sector
 describe spin states in transverse six dimensional space.

The orbifold ${\bf Z}_N$ transformation
 must be a discrete subgroup of SU$(4)_{\rm R}$.
Namely,
\begin{equation}
 \lambda_\alpha \rightarrow e^{2 \pi i a_\alpha / N} \lambda_\alpha
\end{equation}
 with $a_1 +a_2 + a_3 + a_4 = 0 \ {\rm mod} \ N$, and
\begin{equation}
 Z_r \rightarrow e^{- 2 \pi i b_r / N} Z_r
\end{equation}
 with $b_1=a_2+a_3$, $b_2=a_3+a_1$ and $b_3=a_1+a_2$,
 where $Z_r \equiv (X^{2r-1}-iX^{2r})/\sqrt{2}$
 are complexified scalar fields.
In the string world-sheet theory,
 the world-sheet fermion fields transform as
\begin{eqnarray}
 \psi^\mu &\rightarrow& \psi^\mu,
\\
 \psi^{(\pm)r} &\rightarrow& e^{\pm i 2 \pi b_r / N} \psi^{(\pm)r},
\end{eqnarray}
 where
 $\psi^{(\pm)r} \equiv (\psi^{2r+2} \pm i \psi^{2r+3})/\sqrt{2}$.
World-sheet boson fields
 transform in the same way as world-sheet fermion fields.
In case of $a_4=0$ and $b_1+b_2+b_3=0$,
 ${\bf Z}_N \subset$ SU$(3)$ and ${\cal N}=1$ SUSY remains
 ($\lambda_4$ is gaugino).
Chan-Paton indices may be transformed under ${\bf Z}_N$ as
\begin{equation}
 | {\bar i} j \rangle \rightarrow
  (\gamma_3)_{{\bar i}{\bar i}'} 
   | {\bar i}' j' \rangle
  (\gamma_3^{-1})_{j' j},
\end{equation}
 where $\gamma_3 =
  {\rm diag} (I_{n_0}, e^{2 \pi i / N} I_{n_1}, \cdots,
              e^{2 \pi i (N-1) / N} I_{n_{N-1}})$
 with $n_0+n_1+ \cdots +n_{N-1} = n$
 and $I_n$ is the $n \times n$ unit matrix.
By taking ${\bf Z}_N$ invariant states only,
 namely by the ${\bf Z}_N$ orbifold projection,
 the gauge symmetry is broken as
 U$(n) \rightarrow$ U$(n_0) \times$U$(n_1) \times \cdots \times$U$(n_{N-1})$
 and we have the following massless matter fields
 in bi-fundamental representations.
\begin{eqnarray}
 &\mbox{complex scalar fields:}& \qquad
  \sum_{r=1}^3 \sum_{i=0}^{N-1} (n_i, {\bar n}_{i-b_r}),
\\
 &\mbox{Weyl fermion fields:}& \qquad
  \sum_{\alpha=1}^4 \sum_{i=0}^{N-1} (n_i, {\bar n}_{i+a_\alpha}),
\end{eqnarray}
 where $n_i$ and ${\bar n}_i$
 mean fundamental representation and anti-fundamental representation
 of U$(n_i)$, respectively.

Now we take $n=6$
 and consider non-supersymmetric ${\bf Z}_6$ projection of
\begin{eqnarray}
 (a_1,a_1,a_3,a_4) &=& (1,1,1,-3), \\
 (b_1,b_2,b_3) &=& (2,2,2)
\end{eqnarray}
 on world-sheet fields and
\begin{equation}
 (n_0,n_1,\cdots,n_5) = (1,3,2,0,0,0)
\label{D3-solution}
\end{equation}
 on Chan-Paton indices.
Note that
 the transformation on the six-dimensional space is ${\bf Z}_3$,
 therefore, from the geometrical point of view,
 we are considering ${\bf C}^3/{\bf Z_3}$ orbifold singularity.
We have gauge symmetry of U$(3) \times$U$(2) \times$U$(1)$
 and matter of
\begin{eqnarray}
 &\mbox{Higgs doublet fields:} \quad H_r& \qquad
  3 \times (1, 2, -1),
\\
 &\mbox{left-handed quarks:} \quad q_{Lr}& \qquad
  3 \times (3, 2^*, 0),
\\
 &\mbox{right-handed quarks:} \quad u^c{}_{Lr}& \qquad
  3 \times (3^*, 1, +1),
\end{eqnarray}
 where we omit to describe the charges of U$(1)$ factors of
 U$(3)$ and U$(2)$.
The origin of three families is
 the three equivalent complex coordinates
 in six dimensional space (to be compactified)
 under the ${\bf Z}_N$ transformation.
There are Yukawa couplings of
\begin{equation}
 {\cal L}_Y =
  - g \sum_{r,s,t=1,2,3} \epsilon_{rst} {\bar u}_R^r q_L^s H^t + {\rm h.c.}
\end{equation}
 which are originated from the superpotential among three chiral superfields
 in original ${\cal N}=4$ theory ($g$ is the gauge coupling constant).
Higgs fields follow the scalar potential of
\begin{eqnarray}
 V
  &=&
   {{g^2} \over 2}
   \sum_{r,s} \left( H^\dag_r T^a H_r \right)
                      \left( H^\dag_s T^a H_s \right)
   +
  {{g^2} \over 4}
   \sum_{r,s} \left( H^\dag_r H_r \right)
                      \left( H^\dag_s H_s \right)
\nonumber\\
  &=& {{g^2} \over 4}
     \sum_{r,s=1,2,3}
     \left(
      (H^\dag_r H_s)(H^\dag_s H_r)
      +
      (H^\dag_r H_r)(H^\dag_s H_s)
     \right)
\label{potential}
\end{eqnarray}
 which is originated from the D-term potential in the original theory
 ($T^a$ are generator matrices of U$(2)$).
There is no flat direction in this potential,
 and Higgs doublet fields are not D-brane moduli.
 
The gauge anomalies or twisted R-R tadpoles on D3-brane
 are canceled by introducing D7-branes.
We consider the following three types of D7-branes
 (implicitly assuming factorizable toroidal compactifications):
\begin{itemize}
 \item D7${}_1$-brane: world-volume coordinates
       $(x^0,x^1,x^2,x^3), (x^6,x^7), (x^8,x^9),$
 \item D7${}_2$-brane: world-volume coordinates
       $(x^0,x^1,x^2,x^3), (x^4,x^5), (x^8,x^9),$
 \item D7${}_3$-brane: world-volume coordinates
       $(x^0,x^1,x^2,x^3), (x^4,x^5), (x^6,x^7).$
\end{itemize}
Namely, D7${}_r$-brane is point-like in $r$-th complex plane $z_r$.
Assume that $n$ D3-branes and $u$ D7${}_r$-branes
 are on the same point (${\bf C}^3/{\bf Z}_N$ singular point)
 in $z_r$.
We have massless modes of the open string
 whose one edge is on D3-branes and
 another edge is on D7${}_r$-branes.
The boundary conditions of such open string are
 Neumann-Neumann type in four-dimensional space-time directions,
 Dirichlet-Dirichlet type in $z_r$-direction
 and Dirichlet-Neumann type in other directions.
The massless spectrum
 with D3-branes and D7${}_3$-branes, for example,
 before orbifold projection consists of
 two complex scalar fields in NS sector and
 two chiral fermions in R sector.
\begin{eqnarray}
 |s_1,s_2; {\bar i} J \rangle_{\rm NS}
 &\mbox{and}&
 |s_1,s_2; {\bar I} j \rangle_{\rm NS}
 \quad\mbox{with $s_1=s_2=-1/2$},
\\
 |s_3; {\bar i} J \rangle_{\rm R}
 &\mbox{and}&
 |s_3; {\bar I} j \rangle_{\rm R}
 \quad\qquad\mbox{with $s_3=1/2$},
\end{eqnarray}
 where $I,J$ denote Chan-Paton indices on D7${}_3$-brane.
There is a massless U$(u)$ gauge multiplet
 of eight-dimensional ${\cal N}=1$ supersymmetry
 on D7${}_3$-branes.
In case of low string scale (TeV scale),
 the value of the gauge coupling of U$(u)$ becomes very small,
 and U$(u)$ can be considered as a global symmetry.

The ${\bf Z}_N$ transformation of the above open string states are
\begin{eqnarray}
 |s_1,s_2 \rangle_{\rm NS}
  &\rightarrow& e^{- \pi i (b_1 + b_2) / N} |s_1,s_2 \rangle_{\rm NS},
 \\
 |s_3 \rangle_{\rm R}
  &\rightarrow& e^{\pi i b_3 / N} |s_3 \rangle_{\rm R}.
\end{eqnarray}
Note that these phases
 do not belong to ${\bf Z}_N$,
 but they belong to ${\bf Z}_{2N}$. 
Chan-Paton indices may be transformed under ${\bf Z}_N$ as
\begin{eqnarray}
 | {\bar i} J \rangle
 &\rightarrow&
  (\gamma_3)_{{\bar i}{\bar i}'} 
   | {\bar i}' J' \rangle
  (\gamma_{7_3}^{-1})_{J' J},
\\
 | {\bar I} j \rangle
 &\rightarrow&
  (\gamma_{7_3})_{{\bar I}{\bar I}'} 
   | {\bar I}' j' \rangle
  (\gamma_3^{-1})_{j' j},
\end{eqnarray}
 where $\gamma_{7_3} =
  {\rm diag} (I_{u_0}, e^{2 \pi i / N} I_{u_1}, \cdots,
              e^{2 \pi i (N-1) / N} I_{u_{N-1}})$
 with $u_0+u_1+ \cdots +u_{N-1} = u$
 and $I_u$ is the $u \times u$ unit matrix.
Since the phases in state transformation belong to ${\bf Z}_{2N}$,
 we need to set $\gamma_{7_3}$ differently depending on
 whether $b_3$ (or $b_1+b_2=b_3+2a_3$) is even or odd
 to have non-trivial spectra.
In case of even $b_3$,
\begin{equation}
 \gamma_{7_3} =
 {\rm diag} (I_{u_0}, e^{2 \pi i / N} I_{u_1}, \cdots,
             e^{2 \pi i (N-1) / N} I_{u_{N-1}}),
\end{equation}
 and in case of odd $b_3$,
\begin{equation}
 \gamma_{7_3} =
 {\rm diag} (e^{\pi i / N} I_{u_0}, e^{\pi i 3 / N} I_{u_1},
             \cdots,
             e^{\pi i (2N-1) / N} I_{u_{N-1}}),
\end{equation}
 where $u_0+u_1+ \cdots +u_{N-1} = u$
 and $I_u$ is the $u \times u$ unit matrix.
The phases in case of even $b_3$ are
 $e^{2 \pi i m / N}$ with $m =0,1,\cdots,N-1$,
 and the phases in case of odd $b_3$ are
 $e^{2 \pi i m / N} e^{\pi i / N}$ with $m =0,1,\cdots,N-1$.
The resultant massless spectra are as follows.
In case of even $b_3$,
\begin{eqnarray}
 &\mbox{complex scalar fields:}& \qquad
  \sum_{i=0}^{N-1}
  \left[
   (n_i, {\bar u}_{i-{1 \over 2}(b_1+b_2)})
   +
   (u_i, {\bar n}_{i-{1 \over 2}(b_1+b_2)})   
  \right],
\\
 &\mbox{Weyl fermion fields:}& \qquad
  \sum_{i=0}^{N-1}
  \left[
   (n_i, {\bar u}_{i+{1 \over 2}b_3})
   +
   (u_i, {\bar n}_{i+{1 \over 2}b_3})   
  \right],
\end{eqnarray}
 and in case of odd $b_3$,
\begin{eqnarray}
 &\mbox{complex scalar fields:}& \qquad
  \sum_{i=0}^{N-1}
  \left[
   (n_i, {\bar u}_{i-{1 \over 2}(b_1+b_2+1)})
   +
   (u_i, {\bar n}_{i-{1 \over 2}(b_1+b_2-1)})   
  \right],
\\
 &\mbox{Weyl fermion fields:}& \qquad
  \sum_{i=0}^{N-1}
  \left[
   (n_i, {\bar u}_{i+{1 \over 2}(b_3-1)})
   +
   (u_i, {\bar n}_{i+{1 \over 2}(b_3+1)})   
  \right].
\end{eqnarray}
The massless states on D7${}_3$-branes
 get projections in the same way of those on D3-branes.
In case of even $b_3$,
 the resultant massless spectrum consists of
 gauge bosons of
 U$(u_0) \times$U$(u_1) \times \cdots \times$U$(u_{N-1})$
 and
\begin{eqnarray}
 &\mbox{complex scalar fields:}& \qquad
  \sum_{r=1}^3 \sum_{i=0}^{N-1} (u_i, {\bar u}_{i-b_r}),
\\
 &\mbox{Weyl fermion fields:}& \qquad
  \sum_{\alpha=1}^4 \sum_{i=0}^{N-1} (u_i, {\bar u}_{i+a_\alpha}).
\end{eqnarray}
These complex scalar fields can be understood as
 two Wilson lines and one brane modulus.

The twisted R-R tadpole cancellation conditions
 in case of ${\bf C}^3/{\bf Z}_N$ singularity
 and $b_r={\rm even}$ for all $r=1,2,3$ are
\begin{equation}
 \left[
  \prod_{r=1}^3 2 \sin \left( {{\pi k b_r} \over N} \right)
 \right]
 {\rm Tr} \left( (\gamma_3)^k \right)
 +
 \sum_{r=1}^3 2 \sin \left( {{\pi k b_r} \over N} \right)
 {\rm Tr} \left( (\gamma_{7_r})^k \right)
 =0
\end{equation}
 for all $k=1,2,\cdots,N-1$.
In this formula
 we are considering three D7${}_r$-branes ($r=1,2,3$)
 each of which consists $u^r$ D7-branes.
This conditions can be derived
 by calculating open string one-loop vacuum amplitudes
 of D3-D3 sector and D3-D7 sectors.
In the closed string picture after the modular transformation,
 the one-loop vacuum amplitude can be understood
 as the sum of the amplitudes of tree-level propagation
 of closed string modes
 from D3-branes to D3-branes (D3-D3 sector)
 and from D3-branes to D7-branes (D3-D7 sector).
The above conditions are
 the cancellation conditions of the contribution of
 the propagations of massless twisted R-R modes
 between D3-D3 sector and D3-D7 sectors.
It can be shown that
 the gauge symmetry is anomaly free,
 if the twisted R-R tadpole cancellation conditions are satisfied.

Now,
 we consider again the case of $N=6$ and $b_r=2$ for all $r=1,2,3$.
The independent twisted R-R tadpole cancellation conditions are
\begin{equation}
 3 {\rm Tr} \left( (\gamma_3)^k \right)
 +
 \sum_{r=1}^3 {\rm Tr} \left( (\gamma_{7_r})^k \right)
 =0
\label{RR-condition}
\end{equation}
 for $k=1,2$.
We can explicitly write these conditions as
\begin{equation}
 \sum_{j=0}^5 \left( 3 n_j + \sum_{r=1}^3 u^r_j \right)
  (e^{2 \pi i / 6 })^j = 0,
\end{equation}
\begin{equation}
 \sum_{j=0}^2 \left\{
               3 ( n_j + n_{i+3} )
               + \sum_{r-1}^3 ( u^r_j + u^r_{j+3} )
              \right\} (e^{2 \pi i / 6 })^{2j} = 0.
\end{equation}
We find
\begin{equation}
 3 n_j + \sum_{r=1}^3 u^r_j = {\rm const.}
 \quad
 \mbox{for all $j=0,1,\cdots,5$}.
\end{equation}
 is the solution.

Consider only D7${}_3$-brane for simplicity.
The solution in this case is
\begin{equation}
 (u^3_0,u^3_1,\cdots,u^3_5) = (6,0,3,9,9,9)
\label{D7-solution}
\end{equation}
 with Eq.(\ref{D3-solution}).
This solution requires 36 D7${}_3$-branes,
 and the gauge symmetry on the D7${}_3$-brane is
 U$(6) \times$U$(3) \times$U$(9)_1 \times$U$(9)_2 \times$U$(9)_3$.
The massless field contents of this model
 is shown in Table \ref{field-contents}.
The non-anomalous weak hypercharge U$(1)_Y$ with charge $Q_Y$
 is defined as
\begin{equation}
 Q_Y \equiv
  - \left( {{Q_3} \over 3} + {{Q_2} \over 2} + Q_1 \right),
\end{equation}
 where $Q_n$ is the U$(1)$ charges of U$(n)$ on D3-brane.
 
\begin{table}
\begin{tabular}{c|l|c|ccc|ccccc}
 sector & Fermi/Bose & number &
 U$(3)$ & U$(2)$ & U$(1)$ & U$(6)$ & U$(3)$ &
 U$(9)_1$ & U$(9)_2$ & U$(9)_3$ \\
 \hline
 D3-D3 & \quad F \quad $q_L^r$
       & 3 & $3$ & $2^*$ & $0$ & $1$ & $1$ & $1$ & $1$ & $1$ \\
       & \quad F \quad ${\bar u}_L^r$
       & 3 & $3^*$ & $1$ & $+1$ & $1$ & $1$ & $1$ & $1$ & $1$ \\
 \cline{2-11}
       & \quad B \quad $H_u^r$
       & 3 & $1$ & $2$ & $-1$ & $1$ & $1$ & $1$ & $1$ & $1$ \\
 \hline
 D3-D7 & \quad F 
       & 1 & $3$ & $1$ & $0$ & $1$ & $3^*$ & $1$ & $1$ & $1$ \\
       & \quad F \quad $l_L$
       & 1 & $1$ & $2$ & $0$ & $1$ & $1$ & $9^*$ & $1$ & $1$ \\
 \cline{2-11}
       & \quad B \quad $S$
       & 1 & $1$ & $1$ & $+1$ & $1$ & $1$ & $1$ & $9^*$ & $1$ \\
       & \quad B 
       & 1 & $3$ & $1$ & $0$ & $1$ & $1$ & $1$ & $1$ & $9^*$ \\
       & \quad B \quad $H_d$
       & 1 & $1$ & $2$ & $0$ & $6^*$ & $1$ & $1$ & $1$ & $1$ \\
 \hline
 D7-D3 & \quad F \quad ${\bar d}_L$
       & 1 & $3^*$ & $1$ & $0$ & $6$ & $1$ & $1$ & $1$ & $1$ \\
       & \quad F \quad ${\bar e}_L$
       & 1 & $1$ & $1$ & $-1$ & $1$ & $1$ & $1$ & $1$ & $9$ \\
 \cline{2-11} 
       & \quad B 
       & 1 & $1$ & $1$ & $-1$ & $1$ & $3$ & $1$ & $1$ & $1$ \\
       & \quad B \quad $\Phi$ 
       & 1 & $3^*$ & $1$ & $0$ & $1$ & $1$ & $9$ & $1$ & $1$ \\
       & \quad B \quad $H_\nu$
       & 1 & $1$ & $2^*$ & $0$ & $1$ & $1$ & $1$ & $9$ & $1$ \\
 \hline
 D7-D7 & \quad F \quad ${\bar \nu}_L^r$
       & 3 & $1$ & $1$ & $0$ & $1$ & $1$ & $9$ & $9^*$ & $1$ \\
       & \quad $\cdot$ & & & & & & & & & \\
       & \quad $\cdot$ & & & & & & & & & \\
 \hline      
\end{tabular}
\caption{
Massless field contents in the model
 with $6$ D3-branes and $36$ D7${}_3$-branes
 on non-supersymmetric ${\bf C}^3/{\bf Z}_6$ orbifold singularity.
Only one of many fields in D7-D7 sector is listed.
}
\label{field-contents}
\end{table}

There are Yukawa couplings which come from
 the superpotential of the original supersymmetric theory
 due to the open string recombination of
 (D3-D3) $\rightarrow$ (D3-D7${}_3$) + (D7${}_3$-D3):
\begin{equation}
 \Phi q_L^{r=3} l_L, \qquad H_d q_L^{r=3} {\bar d}_L.
\end{equation}
There are scalar quartic interactions of the same origin:
\begin{equation}
 |H_u^{r=3} S |^2, \qquad
 |H_u^{r=3} H_\nu|^2, \qquad
 |S H_\nu|^2. 
\end{equation}
Note that
 only one of three family states on D3-brane
 (only $r=3$ in this case)
 interact with D3-D7${}_3$ and D7${}_3$-D3 states.
This fact can be understood
 by concretely looking the recombination of string states,
 for example
\begin{equation}
 |s_1,s_2 \rangle_{NS} + |s_3 \rangle_R
 \rightarrow |s_1,s_2,s_3 \rangle_R
\end{equation}
 for D3-D7${}_3$ boson ($s_1=s_2=-1/2$)
  + D7${}_3$-D3 fermion ($s_3=1/2$)
  $\rightarrow$ D3-D3 fermion,
 and consider the transformation of each states
 under the ${\bf Z}_N$ transformation.
If we choose the symmetric solution
 among three kinds of D7-branes as
\begin{equation}
 (u^r_0,u^r_1,\cdots,u^r_5) = (2,0,1,3,3,3)
 \quad\mbox{for all $r=1,2,3$}
\end{equation}
 with Eq.(\ref{D3-solution}),
 these interactions become symmetric.
Asymmetric solutions would be more interesting
 than the symmetric solution,
 since three Higgs doublet fields would behave differently
 and obtain different one-loop corrections to their masses.
In this paper
 we concentrate on the asymmetric solution with D7${}_3$-branes only
 and examine the mass of the Higgs doublet field $H_u^3$.

Although
 the twisted R-R tadpoles are canceled out in this model,
 there is no guarantee that
 twisted NS-NS tadpoles are also canceled out simultaneously,
 because of no supersymmetry.
We find that the twisted NS-NS tadpoles are also canceled out
 by explicitly calculating the one-loop open string vacuum amplitude.
After modular transformation,
 the amplitude can be divided into the contributions
 from four sectors.
\begin{eqnarray}
 A_{\rm D3-D3,R-R} &=& \sum_{\gamma=0}^5 A_{\rm D3-D3,R-R}^\gamma,
\\
 A_{\rm D3-D3,NS-NS} &=& \sum_{\gamma=0}^5 A_{\rm D3-D3,NS-NS}^\gamma,
\\
 A_{\rm D7-D7,R-R} &=& \sum_{\gamma=0}^5 A_{\rm D7-D7,R-R}^\gamma,
\\
 A_{\rm D7-D7,NS-NS} &=& \sum_{\gamma=0}^5 A_{\rm D7-D7,NS-NS}^\gamma,
\end{eqnarray}
 where $A_{\rm D3-D3,R-R}$ is the contribution
 from the closed string R-R sector in D3-D3 sector, and so on.
The summation over $\gamma$ comes from
 the ${\bf Z}_6$ projection operator
\begin{equation}
 P_{{\bf Z}_6} = \sum_{\gamma=0}^5 {\hat \alpha}^\gamma
\end{equation}
 in world-sheet theory,
 where ${\hat \alpha}$ is the ${\bf Z}_6$ operator.
In the summation over $\gamma$,
 the amplitudes of $\gamma=0,3$
 are the contributions from the untwisted sector
 and the amplitudes of $\gamma=1,2,4,5$
 are the contributions from twisted sectors.
Since the the projection on the six-dimensional space
 is not by ${\bf Z}_6$ but actually by ${\bf Z}_3$,
 the amplitudes of $\gamma=0,3$
 should be considered as the contribution
 from the untwisted sector.

The asymptotic behavior
 in the limit of long distance propagation of the closed string
 is obtained for D3-D3 sector as follows.
\begin{eqnarray}
 A_{\rm D3-D3,R-R}^\gamma & \rightarrow &
  \int {{ds} \over {2s}}
  \left( {\pi \over s} \right)^2
  {1 \over 2} {1 \over 6}
  {{iV_4} \over {(\sqrt{8 \pi^2 \alpha'})^4}}
  \left( - N_{\rm CP}^\gamma \times 16 \right)
  \quad
  \mbox{for $\gamma=0,3$},
\\ 
 A_{\rm D3-D3,R-R}^\gamma & \rightarrow &
  \int {{ds} \over {2s}}
  {s \over \pi}
  {1 \over 2} {1 \over 6}
  {{iV_4} \over {(\sqrt{8 \pi^2 \alpha'})^4}}
  \left( - N_{\rm CP}^\gamma \times 6 \sqrt{3} \right)
  \quad
  \mbox{for $\gamma \ne 0,3$}
\end{eqnarray}
 for R-R sector and
\begin{eqnarray}
 A_{\rm D3-D3,NS-NS}^\gamma & \rightarrow &
  \int {{ds} \over {2s}}
  \left( {\pi \over s} \right)^2
  {1 \over 2} {1 \over 6}
  {{iV_4} \over {(\sqrt{8 \pi^2 \alpha'})^4}}
  \left( N_{\rm CP}^\gamma \times 16 \right)
  \quad
  \mbox{for $\gamma = 0$},
\\ 
 A_{\rm D3-D3,NS-NS}^\gamma & \rightarrow &
  \int {{ds} \over {2s}}
  \left( {\pi \over s} \right)^2
  {1 \over 2} {1 \over 6}
  {{iV_4} \over {(\sqrt{8 \pi^2 \alpha'})^4}}
  \left( N_{\rm CP}^\gamma \times {1 \over 4} e^s \right)
  \quad
  \mbox{for $\gamma = 3$},  
\\
 A_{\rm D3-D3,NS-NS}^\gamma & \rightarrow &
  \quad
  \mbox{finite (massive modes only) for $\gamma = 1,5$},
\\
 A_{\rm D3-D3,NS-NS}^\gamma & \rightarrow &
  \int {{ds} \over {2s}}
  {s \over \pi}
  {1 \over 2} {1 \over 6}
  {{iV_4} \over {(\sqrt{8 \pi^2 \alpha'})^4}}
  \left( N_{\rm CP}^\gamma \times 6 \sqrt{3} \right)
  \quad
  \mbox{for $\gamma = 2,4$}
\end{eqnarray}
 for NS-NS sector, where
\begin{equation}
 N_{CP}^\gamma
  = ( {\rm Tr} (\gamma_3)^\gamma )
    ( {\rm Tr} (\gamma_3^{-1})^\gamma )
\end{equation}
 is the Chan-Paton factor.
The untwisted sectors and the twisted sectors
 have different asymptotic behaviors.
The R-R sector and NS-NS sector
 also have different asymptotic behaviors.
The amplitudes of $\gamma=0$ R-R and NS-NS untwisted sectors
 and $\gamma=3$ R-R untwisted sectors
 are ``ultraviolet'' ($s \rightarrow \infty$) finite
 due to the power behavior of $(\pi / s)^2$,
 even though massless tadpoles exist in this sector.
The amplitudes of R-R twisted sectors
 and $\gamma=2,4$ NS-NS twisted sectors linearly diverge
 due to the massless twisted R-R and NS-NS tadpoles.
There is a tachyon mode in $\gamma=3$ NS-NS untwisted sector
 which is consistent with the result in Ref.\cite{Font:2002pq}.

The asymptotic form of the amplitudes in D3-D7 sector
 can be obtained as follows.
\begin{eqnarray}
 A_{\rm D3-D7,R-R}^\gamma & = & 0
  \qquad\mbox{amplitudes are zero for $\gamma=0,3$},
\\ 
 A_{\rm D3-D7,R-R}^\gamma & \rightarrow &
  \int {{ds} \over {2s}}
  {s \over \pi}
  {1 \over 2} {1 \over 6}
  {{iV_4} \over {(\sqrt{8 \pi^2 \alpha'})^4}}
  \left( - {\tilde N}_{\rm CP}^\gamma \times 2 \sqrt{3} \right)
  \quad
  \mbox{for $\gamma \ne 0,3$}
\end{eqnarray}
 for R-R sector and
\begin{eqnarray} 
 A_{\rm D3-D7,NS-NS}^\gamma & \rightarrow &
  \quad\mbox{finite (massive modes only) for $\gamma = 0$},
\\ 
 A_{\rm D3-D7,NS-NS}^\gamma & \rightarrow &
  \int {{ds} \over {2s}}
  {1 \over 2} {1 \over 6}
  {{iV_4} \over {(\sqrt{8 \pi^2 \alpha'})^4}}
  \left( - {\tilde N}_{\rm CP}^\gamma \times e^s \right)
  \quad
  \mbox{for $\gamma = 3$},
\\
 A_{\rm D3-D7,NS-NS}^\gamma & \rightarrow &
  \quad
  \mbox{finite (massive modes only) for $\gamma = 1,5$},
\\ 
 A_{\rm D3-D7,NS-NS}^\gamma & \rightarrow &
  \int {{ds} \over {2s}}
  {s \over \pi}
  {1 \over 2} {1 \over 6}
  {{iV_4} \over {(\sqrt{8 \pi^2 \alpha'})^4}}
  \left( {\tilde N}_{\rm CP}^\gamma \times 2 \sqrt{3} \right)
  \quad
  \mbox{for $\gamma = 2,4$}
\end{eqnarray}
 for NS-NS sector, where
\begin{equation}
 {\tilde N}_{CP}^\gamma
  = ( {\rm Tr} (\gamma_3)^\gamma )
    ( {\rm Tr} (\gamma_7^{-1})^\gamma )
\end{equation}
 is the Chan-Paton factor.
There are tadpoles in all R-R twisted sectors
 and $\gamma=2,4$ NS-NS twisted sectors.
There is a tachyon mode in $\gamma=3$ NS-NS untwisted sector.

The twisted R-R tadpoles
 should be canceled out between D3-D3 and D3-D7 sectors.
The condition of the cancellations in the above formulae are
\begin{equation}
 3 N_{\rm CP}^\gamma + {\tilde N}_{\rm CP}^\gamma = 0,
\end{equation}
 and this is nothing but the previously introduced
 twisted R-R tadpole cancellation conditions
 of Eq.(\ref{RR-condition}) for the present case.
It is very interesting that
 the twisted NS-NS tadpoles are also canceled out
 between D3-D3 and D3-D7 sectors with the same conditions.
It is also interesting that
 no tachyon mode exists in twisted sector.
The existence of tachyon modes in twisted sector
 means the instability of the singularity\cite{Adams:2001sv}.
Therefore,
 the present singularity is a stable solution of string theory
 at tree level.

Before closing this section,
 we give a comment on the global definition of the model.
To cancel untwisted R-R tadpoles,
 we have to specify the compact space
 (toroidal orbifolds or toroidal orientifolds, for example),
 and further have to introduce D7-branes, D3-branes
 and their anti-branes.
To project out the tachyon in untwisted sector
 we have to take some special orientifolds
 \cite{Sagnotti:1995ga,Sagnotti:1996qj}.
This step is strongly related to concrete model buildings
 which we do not pursuit in this paper.
The aim of this paper is
 to examine rather model independently
 whether the radiative gauge symmetry breaking is possible or not
 in this kinds of models of D-branes at singularities.

\section{Some calculations of one-loop two point functions}
\label{technique}

We calculate two point function of the gauge boson on D9-brane.
Because of the gauge invariance, the results must be zero.
The following calculations is a demonstration
 to review the technique for superstring one-loop calculation
 in string world-sheet theory.
The technique is directly applicable
 to the one-loop calculation of the masses of the Higgs doublet fields
 on D3-brane at ${\bf C}^3/{\bf Z}_6$ non-supersymmetric singularity.

Before going to the calculation in string theory,
 we stress the importance of the regularization of divergence,
 or the definition of integral,
 by using the example in four-dimensional field theory.
Consider the one-loop correction
 to the mass of U$(1)$ gauge boson by one massive Dirac fermion.
\begin{equation}
 - {\rm tr}
 \int {{d^4 k} \over {(2 \pi)^4 i}}
 g \gamma^\mu
 {1 \over {m-\gamma_\rho k^\rho}}
 g \gamma^\nu
 {1 \over {m-\gamma_\sigma k^\sigma}}
=
 - g^2 g^{\mu\nu}
 \int {{d^4 k} \over {(2 \pi)^4 i}}
 {{4 m^2 - 2 k^2} \over {(m^2-k^2)^2}},  
\end{equation}
 where the external momentum is set to zero.
This integral is divergent.
We use the dimensional regularization,
 which is a gauge invariant regularization,
 to handle this integral.
\begin{eqnarray}
 \int {{d^d k} \over {(2 \pi)^d i}}
  {1 \over {(m^2-k^2)^2}}
 &=& {{\Gamma(2-d/2)} \over {(4 \pi)^{d/2} \Gamma(2)}}
     {1 \over {(m^2)^{2-d/2}}},
\\
 \int {{d^d k} \over {(2 \pi)^d i}}
  {{k^2} \over {(m^2-k^2)^2}}
 &=& - {{\Gamma(2-1-d/2)} \over {(4 \pi)^{d/2} \Gamma(2)}}
     {2 \over {(m^2)^{2-1-d/2}}}
\nonumber\\
 &=& {{-1} \over {2-1-d/2}}
     {{\Gamma(2-d/2)} \over {(4 \pi)^{d/2} \Gamma(2)}}
     {{2m^2} \over {(m^2)^{2-d/2}}},
\end{eqnarray}
 where $d=4-\epsilon$ with $\epsilon \rightarrow 0$ and
 we used the relation $z\Gamma(z)=\Gamma(z+1)$ in the last equality.
We see that this correction to the gauge boson mass vanishes
 by the cancellation between ``quadratically divergent'' part
 and ``logarithmically divergent'' part.

Next, consider the the one-loop correction
 to U$(1)$ gauge boson by one massless Dirac fermion.
\begin{equation}
 - {\rm tr}
 \int {{d^4 k} \over {(2 \pi)^4 i}}
 g \gamma^\mu
 {1 \over {-\gamma_\rho k^\rho}}
 g \gamma^\nu
 {1 \over {-\gamma_\sigma k^\sigma}}
 =
 - 2 g^2 g^{\mu\nu}
 \int {{d^4 k} \over {(2 \pi)^4 i}}
 {1 \over {-k^2}}.
\end{equation}
This quadratically divergent integral
 is regularized or rather defined as follows.
\begin{equation}
 \int {{d^d k} \over {(2 \pi)^d i}} {1 \over {-k^2}}
 = \lim_{\delta \rightarrow 0}
   \int {{d^d k} \over {(2 \pi)^d i}} {1 \over {\delta-k^2}}
 = \lim_{\delta \rightarrow 0}
   {{\Gamma(1-d/2)} \over {(4 \pi)^{d/2} \Gamma(1)}}
   {1 \over {\delta^{1-d/2}}}
 = 0.
\end{equation}
Therefore, this contribution to the gauge boson mass vanishes.
Here, again, the regularization, or definition,
 of the divergent integral is very important to have
 gauge invariant results.
The gauge invariance at one-loop level
 is a non-trivial issue even in four-dimensional field theory.

Now,
 we turn to the one-loop correction in string theory.
Consider open superstring one-loop correction
 to the two point function of massless SU$(N)$ gauge boson states
 on D9-brane.
It is not trivially zero
 like the tree level contribution
 (the disk has three conformal killing vectors
  and the cylinder has one conformal killing vector).
Since we consider SU$(N)$ gauge bosons,
 only the planer diagrams contribute.

We try to apply the formalism
 which is described in the text book by
 Green, Schwarz and Witten\cite{GSW}.
There are two contributions: NS-loop and R-loop,
 which are corresponding to boson and fermion loops, respectively.
\begin{eqnarray}
 A^{\rm NS} &=&
  \int {{d^{10}k} \over {(2\pi)^{10} i}}
  {\rm tr} \left\{ \Delta \ V(1) \ \Delta \ V(1) \ P_{\rm GSO}
           \right\},
\\
 A^{\rm R} &=&
  - \int {{d^{10}k} \over {(2\pi)^{10} i}}
  {\rm tr} \left\{ S \ W(1) \ S \ W(1) \ P_{\rm GSO} \right\},
\end{eqnarray}
 where $P_{\rm GSO}$ is the GSO projection operator,
\begin{eqnarray}
 \Delta &=& \int_0^1 x^{L_0 -1} dx,
\\
 S &=& G_0 \int_0^1 x^{L_0-1} dx = G_0 \Delta
\end{eqnarray}
 are propagator operators
 ($L_0$ and $G_0$ are generators in super-Virasoro algebra),
\begin{eqnarray}
 V(x) &=&
  {{g_O} \over \sqrt{2 \alpha'}} e_\mu i {\dot X}^\mu,
  \qquad
  {\dot X}^\mu \equiv x {d \over {dx}} X^\mu(x),
\\
 W(x) &=&
  g_O \sqrt{2} e_\mu \sqrt{x} \psi^\mu(x)
\end{eqnarray}
 are vertex operators with zero external momentum,
 $e_\mu$ is the polarization vector with Chan-Paton indices,
 and $g_O$ is the open string coupling.
The argument $z=e^{-i(\sigma_1+i\sigma_2)}$
 of the world-sheet fields, $X^\mu(z)$ and $\psi^\mu(z)$,
 is now real: $z=x$.

First, consider NS-loop.
The straightforward calculation gives the following result.
\begin{eqnarray}
 A^{\rm NS} &=&
 g_O^2 {\rm tr}(e_\mu e_\nu) \eta^{\mu\nu}
 {1 \over {(2 \pi)^{10}}}
 \int_0^1 {{d\rho_1} \over {\rho_1}} {{d\rho_2} \over {\rho_2}}
          \theta(\rho_1 - \rho_2)
\nonumber\\
 & \times &
 \left( \sqrt{{\pi \over {-\alpha' \ln \rho_2}}} \right)^{10}
 \left\{
  {1 \over {- \ln \rho_2}}
  +
  \sum_{l=1}^\infty l
  {{(\rho_2/\rho_1)^l+\rho_2^l/(\rho_2/\rho_1)^l}
   \over
   {1-\rho_2^l}}   
 \right\}
\nonumber\\
 & \times &
 {1 \over {(\eta(\tau_2))^8}}
 \cdot {1 \over 2}
 \left\{
  \left( {{\theta_3(\tau_2)} \over {\eta(\tau_2)}} \right)^4
  -
  \left( {{\theta_4(\tau_2)} \over {\eta(\tau_2)}} \right)^4
 \right\},
\end{eqnarray}
 where $\tau_2$ is defined by $\rho_2 = \exp(2 \pi i \tau_2)$.
The factor in the first curly brackets is proportional to
 the correlation function
 $\langle {\dot X}^\mu(\rho_1) {\dot X}^\nu(\rho_2) \rangle$
 on the boundary circle.
Consider the integration by $\rho_1$:
\begin{equation}
 I^{\rm NS} =
 \int_0^1 {{d\rho_1} \over {\rho_1}} \theta(\rho_1 - \rho_2)
 \left\{
  {1 \over {- \ln \rho_2}}
  +
  \sum_{l=1}^\infty l
  {{(\rho_2/\rho_1)^l+\rho_2^l/(\rho_2/\rho_1)^l}
   \over
   {1-\rho_2^l}}   
 \right\}
 \end{equation}
The first term becomes
\begin{equation}
 \int_0^1 {{d\rho_1} \over {\rho_1}} \theta(\rho_1 - \rho_2)
 {1 \over {- \ln \rho_2}}
 = \int_{\rho_2}^1 {{d\rho_1} \over {\rho_1}}
   {1 \over {- \ln \rho_2}}
 = 1.
\end{equation}
The second term is a divergent quantity,
 and we have to define this quantity by some regularization.
Consider the following regularization:
\begin{eqnarray}
 \int_{\rho_2}^1 {{d\rho_1} \over {\rho_1}}
 \sum_{l=1}^\infty l
  {{(\rho_2/\rho_1)^l+\rho_2^l/(\rho_2/\rho_1)^l}
   \over
   {1-\rho_2^l}}
 &\rightarrow&
  \sum_{l=1}^\infty l
  \int_{\rho_2}^1 {{d\rho_1} \over {\rho_1}}
  {{(\rho_2/\rho_1)^l+\rho_2^l/(\rho_2/\rho_1)^l}
   \over
   {1-\rho_2^l}}
\nonumber\\
 &=&
 2 \sum_{l=1}^\infty 1
\nonumber\\
 &\rightarrow&
 2 \lim_{z \rightarrow 0} \sum_{l=1}^\infty {1 \over {l^z}}
 = 2 \lim_{z \rightarrow 0} \zeta(z) = -1, 
\end{eqnarray}
 where
 $\zeta(z)$ is the Riemann zeta function and
 $\zeta(0)$ is defined as $-1/2$ by the analytic continuation.
Therefore,
 $I^{\rm NS}=0$ and $A^{\rm NS}=0$
 as required by gauge invariance.

Next, consider R-loop.
The straightforward calculation
 (using the above zeta function regularization in part)
 gives the following result.
\begin{eqnarray}
 A^{\rm R} &=&
 - g_O^2 {\rm tr}(e_\mu e_\nu) \eta^{\mu\nu}
 {1 \over {(2 \pi)^{10}}}
 \int_0^1 {{d\rho_1} \over {\rho_1}} {{d\rho_2} \over {\rho_2}}
          \theta(\rho_1 - \rho_2)
\nonumber\\
 & \times &
 \left( \sqrt{{\pi \over {-\alpha' \ln \rho_2}}} \right)^{10}
 \left\{
  {1 \over {- \ln \rho_2}}
  +
  2 \sum_{l=1}^\infty l \rho_2^l
  {{(\rho_2/\rho_1)^l+(\rho_1/\rho_2)^l}
   \over
   {(1-\rho_2^l)(1+\rho_2^l)}}   
 \right\}
\nonumber\\
 & \times &
 {1 \over {(\eta(\tau_2))^8}}
 \cdot {1 \over 2}
 \left( {{\theta_2(\tau_2)} \over {\eta(\tau_2)}} \right)^4.
\end{eqnarray}
The part
 which is proportional to $1/\ln \rho_2$ in the integrant
 is canceled out in $A^{\rm NS}+A^{\rm R}$
 due to the identity
 $(\theta_3)^4 - (\theta_4)^4 - (\theta_2)^4 = 0$.
This is the result of supersymmetry.
Consider the integration by $\rho_1$.
\begin{equation}
 I^{\rm R} =
  \int_0^1 {{d\rho_1} \over {\rho_1}} \theta(\rho_1 - \rho_2)
  \left\{
  {1 \over {- \ln \rho_2}}
  +
  2 \sum_{l=1}^\infty l \rho_2^l
  {{(\rho_2/\rho_1)^l+(\rho_1/\rho_2)^l}
   \over
   {(1-\rho_2^l)(1+\rho_2^l)}}
  \right\}.
\end{equation}
The second term is a divergent quantity,
 and again we have to define this quantity by some regularization.
Consider the same regularization procedure in NS-loop calculation.
\begin{eqnarray}
 \int_{\rho_2}^1 {{d\rho_1} \over {\rho_1}}
  2 \sum_{l=1}^\infty l \rho_2^l
  {{(\rho_2/\rho_1)^l+(\rho_1/\rho_2)^l}
   \over
   {(1-\rho_2^l)(1+\rho_2^l)}}
 &\rightarrow&
  2 \sum_{l=1}^\infty l
  \int_{\rho_2}^1 {{d\rho_1} \over {\rho_1}}
  \rho_2^l
  {{(\rho_2/\rho_1)^l+(\rho_1/\rho_2)^l}
   \over
   {(1-\rho_2^l)(1+\rho_2^l)}}
\nonumber\\
 &=&
 2 \sum_{l=1}^\infty 1
\nonumber\\
 &\rightarrow&
 2 \lim_{z \rightarrow 0} \sum_{l=1}^\infty {1 \over {l^z}}
 = 2 \lim_{z \rightarrow 0} \zeta(z) = -1.
\end{eqnarray}
Therefore,
 $I^{\rm R}=0$ and $A^{\rm R}=0$
 as required by gauge invariance.

In the next section,
 we use the same definition of the divergent integrals
 in the calculation of mass squared of Higgs doublet field.

\section{One-loop correction to the Higgs mass}
\label{results}

The two point function of the Higgs doublet field
 can be calculated using the same technique
 reviewed in the previous section.
We calculate one-loop two points function of $H_u^3$,
 one of three Higgs doublet fields,
 with zero external momentum.
There are two contributions from D3-D3 and D3-D7 sectors.
Only the planer diagrams contribute
 because of the conservation of Chan-Paton charges
 (U$(2) \times$U$(1)$ charges).
 
In D3-D3 sector
 there are two planer diagrams corresponding to
 which boundary of annulus two open string vertex operators
 attach to.
It is enough to calculate one of them,
 because the difference appears only in Chan-Paton factors:
 they are complex conjugate with each other.
The contribution of one planar diagram in D3-D3 sector
 is given by
\begin{equation}
 A_{\rm D3-D3}=A^{\rm NS}_{\rm D3-D3}+A^{\rm R}_{\rm D3-D3}
\end{equation}
 with
\begin{eqnarray}
 A^{\rm NS}_{\rm D3-D3} &=&
  \int {{d^4 k} \over {(2\pi)^4 i}}
  {\rm tr}
   \left\{ \Delta \ V^{(-)}(1) \ \Delta \ V^{(+)}(1)
           \ P_{{\bf Z}_6} \ P_{\rm GSO} \right\}
  + (- \leftrightarrow +),
\label{NS-loop}
\\
 A^{\rm R}_{\rm D3-D3} &=&
  - \int {{d^4 k} \over {(2\pi)^4 i}}
  {\rm tr}
   \left\{ S \ W^{(-)}(1) \ S \ W^{(+)}(1)
           \ P_{{\bf Z}_6} \ P_{\rm GSO} \right\}
  + (- \leftrightarrow +),
\label{R-loop}
\end{eqnarray}
 where
\begin{eqnarray}
 V^{(+)}(x) &=&
  {{g_O} \over \sqrt{2 \alpha'}} u i {\dot X}^{(+)}(x)
 \quad\mbox{and}\quad
 V^{(-)}(x) =
  {{g_O} \over \sqrt{2 \alpha'}} u^\dag i {\dot X}^{(-)}(x),
\\
 W^{(+)}(x) &=&
  g_O \sqrt{2} u \sqrt{x} \psi^{(+)}(x)
 \quad\mbox{and}\quad
 W^{(-)}(x) =
  g_O \sqrt{2} u^\dag \sqrt{x} \psi^{(-)}(x)
\end{eqnarray}
 are vertex operators with zero external momentum
 with
\begin{equation}
 X^{(\pm)} \equiv {1 \over \sqrt{2}} ( X^8 \pm i X^9)
 \quad\mbox{and}\quad
 \psi^{(\pm)} \equiv {1 \over \sqrt{2}} ( \psi^8 \pm i \psi^9).
\end{equation}
The factor $u$ carries Chan-Paton indices.
A special care is required
 to correctly include the action of $P_{{\bf Z}_6}$
 on Chan-Paton indices.
The open string coupling constant $g_O$
 is translated to the dimensionless coupling
 (gauge coupling on D3-brane) by $g = g_O/\sqrt{\alpha'}$.
There are two contributions, NS-loop and R-loop,
 which are corresponding to boson and fermion loops,
 respectively.
The amplitude $A_{\rm D3-D3}$ can be understood as mass squared,
 and the contribution of NS-loop (R-loop) is positive (negative).
The boundary condition of the open string
 is Neumann-Neumann type in the direction in D3-brane world-volume,
 and Dirichlet-Dirichlet type in the directions
 of transverse six-dimensional space.
No space-time momentum in the transverse six-dimensional space
 is allowed due to the Dirichlet-Dirichlet boundary condition.

The straightforward calculation results
\begin{eqnarray}
 A^{\rm NS}_{\rm D3-D3} &=&
  {{g_O^2} \over {(2\pi)^4}}
  {1 \over 6} \sum_{\gamma=0}^5 N_{\rm D3}^\gamma
  \int_0^1 {{d \rho} \over {\rho}}
  \left( \sqrt{{\pi \over {- \alpha' \ln \rho}}} \right)^4
\nonumber\\
&&
 \times
 2 \Re
 \left\{
  -1
  + (1-e^{2 \pi i \gamma /3})
    \left( {1 \over 2} -
    2 i \sum_{n=1}^\infty
         \sin ({{2 \pi n \gamma} \over 3})
         {{\rho^n} \over {1-\rho^n}}
    \right)
 \right\}
\nonumber\\
&&
 \times
 {1 \over 2}
 \left[
  {{\theta_3(\tau)} \over {\eta(\tau)^3}}
  \left(
   {{\theta
     \left[
      \begin{array}{c}
       0 \\ -\gamma/3
      \end{array}
     \right](0|\tau)
    }
    \over
    {
     \theta
     \left[
      \begin{array}{c}
       1/2 \\ \gamma/3
      \end{array}
     \right](-1/2|\tau)
     {\Big /} 2 \sin (\pi \gamma/3)
    }
   }
  \right)^3
 \right.
\nonumber\\
&& \qquad
 \left.
 -
  {{\theta_4(\tau)} \over {\eta(\tau)^3}}
  \left(
   {{\theta
     \left[
      \begin{array}{c}
       0 \\ 1/2-\gamma/3
      \end{array}
     \right](0|\tau)
    }
    \over
    {
     \theta
     \left[
      \begin{array}{c}
       1/2 \\ \gamma/3
      \end{array}
     \right](-1/2|\tau)
     {\Big /} 2 \sin (\pi \gamma/3)
    }
   }
  \right)^3
 \right]
\end{eqnarray}
 and
\begin{eqnarray}
 A^{\rm R}_{\rm D3-D3} &=&
  - {{g_O^2} \over {(2\pi)^4}}
  {1 \over 6} \sum_{\gamma=0}^5 N_{\rm D3}^\gamma
  \int_0^1 {{d \rho} \over {\rho}}
  \left( \sqrt{{\pi \over {- \alpha' \ln \rho}}} \right)^4
\nonumber\\
&&
 \times
 2 \Re
 \left\{
  -1
  - (1-e^{2 \pi i \gamma /3})
    \left(
    4 i \sum_{n=1}^\infty
         {{1-(-1)^n} \over 2}
         \sin ({{2 \pi n \gamma} \over 3})
         {{\rho^n} \over {1-\rho^n}}
    \right)
 \right\}
\nonumber\\
&&
 \times
 {1 \over 2}
  {{\theta_2(\tau)} \over {\eta(\tau)^3}}
  \left(
   {{\theta
     \left[
      \begin{array}{c}
       1/2 \\ -\gamma/3
      \end{array}
     \right](0|\tau)
    }
    \over
    {
     \theta
     \left[
      \begin{array}{c}
       1/2 \\ \gamma/3
      \end{array}
     \right](-1/2|\tau)
     {\Big /} 2 \sin (\pi \gamma/3)
    }
   }
  \right)^3,
\end{eqnarray}
 where $\tau = it$ with $\rho=e^{2 \pi i \tau}=e^{-2 \pi t}$,
\begin{equation}
 N_{\rm D3}^\gamma \equiv 2 {\rm tr} (\gamma_3^{-1})
\end{equation}
 and
\begin{eqnarray}
 \theta
  \left[
   \begin{array}{c}
    \alpha \\ \beta
   \end{array}
  \right](z|\tau)
  &\equiv&
  e^{2 \pi i \alpha (z + \beta)}
  q^{\alpha^2/2}
  \prod_{n=1}^\infty (1-q^n)
\nonumber\\
&&
  \times
  \prod_{m=1}^\infty (1+q^{m+\alpha-1/2}e^{2 \pi i (z+\beta)})
                     (1+q^{m-\alpha-1/2}e^{- 2 \pi i (z+\beta)})
\end{eqnarray}
 is the generalized theta function ($q \equiv \exp(2\pi i \tau)$)
 \cite{Angelantonj:2002ct}.
Three well-known theta functions
 are given by some special cases
 of this generalized theta function as
\begin{equation}
 \theta_2(\tau) =
 \theta
  \left[
   \begin{array}{c}
    1/2 \\ 0
   \end{array}
  \right](0|\tau),
\quad
 \theta_3(\tau) =
 \theta
  \left[
   \begin{array}{c}
    0 \\ 0
   \end{array}
  \right](0|\tau),
\quad
 \theta_4(\tau) =
 \theta
  \left[
   \begin{array}{c}
    0 \\ 1/2
   \end{array}
  \right](0|\tau).
\end{equation}

There are also two planer diagrams in D3-D7 sector,
 because there are two ways of assigning Chan-Paton indices
 to the open string in a loop.
It is enough to calculate one of them,
 because the difference appears only in Chan-Paton factors:
 they are again complex conjugate with each other.
The contribution of one planar diagram in D3-D7 sector
 is given by
\begin{equation}
 A_{\rm D3-D7}=A^{\rm NS}_{\rm D3-D7}+A^{\rm R}_{\rm D3-D7}
\end{equation}
 with exactly the same form in Eqs.(\ref{NS-loop}) and (\ref{R-loop})
 for $A^{\rm NS}_{\rm D3-D7}$ and $A^{\rm R}_{\rm D3-D7}$,
 respectively.
The main difference from D3-D3 sector
 is the boundary condition of the open string.
We have to take Neumann-Neumann boundary condition
 for the directions in D3-brane world-volume,
 Dirichlet-Dirichlet boundary condition
 for the transverse directions of D7-brane,
 and Dirichlet-Neumann boundary condition
 for the other directions.

The straightforward calculation gives the following results.
\begin{eqnarray}
 A^{\rm NS}_{\rm D3-D7} &=&
  {{g_O^2} \over {(2\pi)^4}}
  {1 \over 6} \sum_{\gamma=0}^5 N_{\rm D7}^\gamma
  \int_0^1 {{d \rho} \over {\rho}}
  \left( \sqrt{{\pi \over {- \alpha' \ln \rho}}} \right)^4
\\
&&
 \times
 2 \Re
 \left\{
  -1
  + (1-e^{2 \pi i \gamma /3})
    \left( {1 \over 2} -
    2 i \sum_{n=1}^\infty
         \sin ({{2 \pi n \gamma} \over 3})
         {{\rho^n} \over {1-\rho^n}}
    \right)
 \right\}
\nonumber\\
&&
 \times
 {1 \over 2}
 \left[
  {{\theta_3(\tau)} \over {\eta(\tau)^3}}
   {{\theta
     \left[
      \begin{array}{c}
       0 \\ -\gamma/3
      \end{array}
     \right](0|\tau)
    }
    \over
    {
     \theta
     \left[
      \begin{array}{c}
       1/2 \\ \gamma/3
      \end{array}
     \right](-1/2|\tau)
     {\Big /} 2 \sin (\pi \gamma/3)
    }
   }
  \left(
   {{\theta
     \left[
      \begin{array}{c}
       -1/2 \\ -\gamma/3
      \end{array}
     \right](0|\tau)
    }
    \over
    {
     \theta
     \left[
      \begin{array}{c}
       0 \\ \gamma/3
      \end{array}
     \right](-1/2|\tau)
    }
   }
  \right)^2
 \right.
\nonumber\\
&& \qquad
 \left.
 -
  {{\theta_4(\tau)} \over {\eta(\tau)^3}}
   {{\theta
     \left[
      \begin{array}{c}
       0 \\ 1/2-\gamma/3
      \end{array}
     \right](0|\tau)
    }
    \over
    {
     \theta
     \left[
      \begin{array}{c}
       1/2 \\ \gamma/3
      \end{array}
     \right](-1/2|\tau)
     {\Big /} 2 \sin (\pi \gamma/3)
    }
   }
  \left(
   {{\theta
     \left[
      \begin{array}{c}
       -1/2 \\ 1/2-\gamma/3
      \end{array}
     \right](0|\tau)
    }
    \over
    {
     \theta
     \left[
      \begin{array}{c}
       0 \\ \gamma/3
      \end{array}
     \right](-1/2|\tau)
    }
   }
  \right)^2
 \right]
\nonumber
\end{eqnarray}
 and
\begin{eqnarray}
 A^{\rm R}_{\rm D3-D7} &=&
  - {{g_O^2} \over {(2\pi)^4}}
  {1 \over 6} \sum_{\gamma=0}^5 N_{\rm D7}^\gamma
  \int_0^1 {{d \rho} \over {\rho}}
  \left( \sqrt{{\pi \over {- \alpha' \ln \rho}}} \right)^4
\\
&&
 \times
 2 \Re
 \left\{
  -1
  - (1-e^{2 \pi i \gamma /3})
    \left(
    4 i \sum_{n=1}^\infty
         {{1-(-1)^n} \over 2}
         \sin ({{2 \pi n \gamma} \over 3})
         {{\rho^n} \over {1-\rho^n}}
    \right)
 \right\}
\nonumber\\
&&
 \times
 {1 \over 2}
  {{\theta_2(\tau)} \over {\eta(\tau)^3}}
   {{\theta
     \left[
      \begin{array}{c}
       1/2 \\ -\gamma/3
      \end{array}
     \right](0|\tau)
    }
    \over
    {
     \theta
     \left[
      \begin{array}{c}
       1/2 \\ \gamma/3
      \end{array}
     \right](-1/2|\tau)
     {\Big /} 2 \sin (\pi \gamma/3)
    }
   }
  \left(
   {{\theta
     \left[
      \begin{array}{c}
       0 \\ -\gamma/3
      \end{array}
     \right](0|\tau)
    }
    \over
    {
     \theta
     \left[
      \begin{array}{c}
       0 \\ \gamma/3
      \end{array}
     \right](-1/2|\tau)
    }
   }
  \right)^2,
\nonumber
\end{eqnarray}
 where
\begin{equation}
 N_{\rm D7}^\gamma \equiv 2 {\rm tr} (\gamma_7^{-1}).
\end{equation}
Twisted R-R tadpole cancellation conditions
 are encoded in the relations of
\begin{equation}
 N_{\rm D7}^\gamma = - 3 N_{\rm D3}^\gamma.
\end{equation}

The one-loop correction to the Higgs mass squared
 is obtained by combining above results as follows.
\begin{eqnarray}
 m^2 &=&
 {{(g_O/\sqrt{\alpha'})^2} \over {16 \pi^2}}
 {2 \over {\alpha'}} \int_0^1 {{d\rho} \over {\rho (\ln\rho)^2}}
\label{m^2}\\
&& \times
 \left\{
  - {1 \over 6} \sum_{\gamma=0}^5
    \Re(N_{\rm D3}^\gamma)
    \left( Z_{{\rm NS}, \beta=0}^\gamma 
           - Z_{{\rm NS}, \beta=1}^\gamma \right)
 \right.
\nonumber\\
&& \qquad
  + {1 \over 6} \sum_{\gamma=0}^5
    \Re(N_{\rm D3}^\gamma)
    \Re({{1-e^{2 \pi i \gamma / 3}} \over 2})
    \left( Z_{{\rm NS}, \beta=0}^\gamma 
           - Z_{{\rm NS}, \beta=1}^\gamma \right)
\nonumber\\
&& \qquad
  - {1 \over 6} \sum_{\gamma=0}^5
    \Re(N_{\rm D3}^\gamma)
    \Re(i(1-e^{2 \pi i \gamma / 3}))
      2 \sum_{n=1}^\infty \sin ({{2 \pi n \gamma} \over 3})
      {{\rho^n} \over {1-\rho^n}}
    \left( Z_{{\rm NS}, \beta=0}^\gamma 
           - Z_{{\rm NS}, \beta=1}^\gamma \right)
\nonumber\\
&& \qquad
  + {1 \over 6} \sum_{\gamma=0}^5
    \Re(N_{\rm D3}^\gamma)
    Z_{{\rm R},\beta=0}
\nonumber\\
&& \qquad
 \left.
  + {1 \over 6} \sum_{\gamma=0}^5
    \Re(N_{\rm D3}^\gamma)
    \Re(i(1-e^{2 \pi i \gamma / 3}))
      2 \sum_{n=1}^\infty (1-(-1)^n)
      \sin ({{2 \pi n \gamma} \over 3})
      {{\rho^n} \over {1-\rho^n}}
    Z_{{\rm R},\beta=0}
 \right\},
\nonumber
\end{eqnarray}
 where the definition of
  $Z_{{\rm NS}, \beta=0}$, $Z_{{\rm NS}, \beta=1}$
  and $Z_{{\rm R}, \beta=0}$ are given in Appendix.
The first three terms
 in the curly brackets in the above equation
 describe the one-loop contribution of bosonic modes,
 and the last two terms describe the contribution of fermionic modes.
It is expected that
 the boson loop contributes positively
 and the fermion loop contributes negatively.
As explained in section \ref{intro},
 we consider only twisted sectors, $\gamma=1,2,4,5$,
 because of the model (compactification) dependence
 of the untwisted sector, $\gamma=0,3$.

Since present ${\bf Z}_6$ transformation
 acts as ${\bf Z}_3$ transformation in bosonic sector
 (present orbifold singularity
  is geometrically ${\bf C}^3/{\bf Z}_3$),
 the factor
 $Z_{{\rm NS}, \beta=0}^\gamma - Z_{{\rm NS}, \beta=1}^\gamma$
 is the function of $\exp(2 \pi i \gamma / 3)$
 and its complex conjugate as the function of $\gamma$.
Therefore,
 the values in case of $\gamma=1$ and $\gamma=4$ equal,
 and the values in case of $\gamma=2$ and $\gamma=5$ equal.
On the other hand,
 all the three coefficients in this factor in Eq.(\ref{m^2})
 give the values of the same magnitude with opposite sign
 for each pair of $\gamma=1,4$ and $\gamma=2,5$.
Consequently,
 all the boson loop contributions of twisted sectors
 are canceled out in this model.
This is not the case in fermion loop sector,
 since the fermion states fully transform under ${\bf Z}_6$.
The first term of the fermion loop contributions in Eq.(\ref{m^2})
 diverges at $\rho \rightarrow 0$,
 which corresponds to the effect of massless physical states
 in the loop.
Since we do not follow the ``$i \epsilon$-prescription''
 in the definition of the propagator operators,
 we have divergence of the type of $\ln(m)$ with $m \rightarrow 0$
 instead of some imaginary part.
We neglect this divergent contribution
 and consider only the last term of Eq.(\ref{m^2}).
Which should give negative contribution to Higgs mass squared,
 since it is a fermion loop contribution.

We can make an order estimate
 by moving to the closed string picture
 and neglecting the exponentially suppressed higher-order terms
 in the integrant. 
The result is
\begin{equation}
 m^2 \sim
 -
 {{g^2} \over {16 \pi^2}}
 {2 \over {\alpha'}}
 {{36\sqrt{3}} \over \pi}
 \int_0^\infty {{ds} \over s}
 {{e^{- 2 \pi^2 / s}} \over {1-e^{- 2 \pi^2 / s}}}
 e^{-s/3},
\end{equation}
 where $s \equiv \pi/t$ and $g=g_O/\sqrt{\alpha'}$.
We can have finite result
 due to twisted tadpole cancellations in
 both R-R and NS-NS sectors.
The order of the ``electroweak scale'' is numerically obtained as
\begin{equation}
 v \sim \sqrt{-m^2/g^2} \simeq 10^{-2} \alpha'^{-1/2},
\end{equation}
 where we used that the Higgs quartic coupling
 is given by the gauge coupling, Eq.(\ref{potential}).
This is consistent with the standard understanding
 that the scale is given by string scale
 with one-loop suppression factor.
This result suggests that
 the radiative gauge symmetry breaking on D-branes is possible
 in this type of models of D-branes
 at non-supersymmetric orbifold singularities.

At the end of this paper,
 we would like to give some comments
 on the global definition of the model.
To have some consistent models
 with the tadpole cancellation in untwisted sector,
 we have to specify some concrete six-dimensional compact space,
 which is an orbifold or orientifold.
By setting
 appropriate D3-branes and D7-branes
 at one of some orbifold fixed points in that compact space,
 it is possible to have standard model like massless spectrum.
The tadpoles in untwisted sector can be canceled out by putting
 some D3-branes, D7-branes, anti-D3-branes and anti-D7-branes,
 at some appropriate other fixed points
 in that compact space.
 (some examples are given in Ref.\cite{Aldazabal:2000sa}).
Since fixed points are specially separated,
 no new massless mode with the gauge charge of our D3-branes
 appears.
If the distance between fixed points
 can be taken very large in compare with the string scale
 (the size of the compact space
  must be large in TeV string scenario anyway),
 the contribution to the Higgs mass squared
 through the one-loop effect by these massive open string modes
 would be small.
The tachyon modes in untwisted sector
 may be projected out by taking certain orientifold projections
 (see Ref.\cite{Angelantonj:2002ct}
  for open descendants of the type 0B model, for example).
Therefore,
 it would be possible to have some consistent systems
 without tadpoles and tachyon modes in untwisted sector
 with appropriate massless modes on D3-branes.
It may be interesting to estimate
 how large the one-loop contribution of untwisted sector
 to the Higgs mass squared in these globally well-defined models.
(In the present model
 only $\gamma=0$ untwisted sector contributes in Eq.(\ref{m^2}),
 since $\Re (N_{\rm D3}^{\gamma=3}) = 0$.) 
It would be very interesting
 to explorer realistic models in this direction.

\section*{Acknowledgment}

I would like to thank E.~Dudas and A.~Sagnotti
 for helpful comments.

\section*{Appendix}

The definition of the functions
 $Z_{{\rm NS}, \beta=0}$, $Z_{{\rm NS}, \beta=1}$
 and $Z_{{\rm R}, \beta=0}$ in Eq.(\ref{m^2})
 is given as follows.

\begin{eqnarray}
 Z_{{\rm NS}, \beta=0}^\gamma &=&
  {{\theta_3(\tau)} \over {\eta(\tau)^3}}
   {{\theta
     \left[
      \begin{array}{c}
       0 \\ -\gamma/3
      \end{array}
     \right](0|\tau)
    }
    \over
    {
     \theta
     \left[
      \begin{array}{c}
       1/2 \\ \gamma/3
      \end{array}
     \right](-1/2|\tau)
     {\Big /} 2 \sin (\pi \gamma/3)
    }
   }
\\
&& \times
 \left\{
  \left(
   {{\theta
     \left[
      \begin{array}{c}
       0 \\ -\gamma/3
      \end{array}
     \right](0|\tau)
    }
    \over
    {
     \theta
     \left[
      \begin{array}{c}
       1/2 \\ \gamma/3
      \end{array}
     \right](-1/2|\tau)
     {\Big /} 2 \sin (\pi \gamma/3)
    }
   }
  \right)^2
  - 3
  \left(
   {{\theta
     \left[
      \begin{array}{c}
       -1/2 \\ -\gamma/3
      \end{array}
     \right](0|\tau)
    }
    \over
    {
     \theta
     \left[
      \begin{array}{c}
       0 \\ \gamma/3
      \end{array}
     \right](-1/2|\tau)
    }
   }
  \right)^2
 \right\},
\nonumber\\
 Z_{{\rm NS}, \beta=1}^\gamma &=&
  {{\theta_4(\tau)} \over {\eta(\tau)^3}}
   {{\theta
     \left[
      \begin{array}{c}
       0 \\ 1/2-\gamma/3
      \end{array}
     \right](0|\tau)
    }
    \over
    {
     \theta
     \left[
      \begin{array}{c}
       1/2 \\ \gamma/3
      \end{array}
     \right](-1/2|\tau)
     {\Big /} 2 \sin (\pi \gamma/3)
    }
   }
\\
&& \times
 \left\{
  \left(
   {{\theta
     \left[
      \begin{array}{c}
       0 \\ 1/2-\gamma/3
      \end{array}
     \right](0|\tau)
    }
    \over
    {
     \theta
     \left[
      \begin{array}{c}
       1/2 \\ \gamma/3
      \end{array}
     \right](-1/2|\tau)
     {\Big /} 2 \sin (\pi \gamma/3)
    }
   }
  \right)^2
  - 3
  \left(
   {{\theta
     \left[
      \begin{array}{c}
       -1/2 \\ 1/2-\gamma/3
      \end{array}
     \right](0|\tau)
    }
    \over
    {
     \theta
     \left[
      \begin{array}{c}
       0 \\ \gamma/3
      \end{array}
     \right](-1/2|\tau)
    }
   }
  \right)^2
 \right\},
\nonumber\\
 Z_{{\rm R},\beta=0} &=&
  {{\theta_2(\tau)} \over {\eta(\tau)^3}}
   {{\theta
     \left[
      \begin{array}{c}
       1/2 \\ -\gamma/3
      \end{array}
     \right](0|\tau)
    }
    \over
    {
     \theta
     \left[
      \begin{array}{c}
       1/2 \\ \gamma/3
      \end{array}
     \right](-1/2|\tau)
     {\Big /} 2 \sin (\pi \gamma/3)
    }
   }
\\
&& \times
 \left\{
  \left(
   {{\theta
     \left[
      \begin{array}{c}
       1/2 \\ -\gamma/3
      \end{array}
     \right](0|\tau)
    }
    \over
    {
     \theta
     \left[
      \begin{array}{c}
       1/2 \\ \gamma/3
      \end{array}
     \right](-1/2|\tau)
     {\Big /} 2 \sin (\pi \gamma/3)
    }
   }
  \right)^2
  - 3
  \left(
   {{\theta
     \left[
      \begin{array}{c}
       0 \\ -\gamma/3
      \end{array}
     \right](0|\tau)
    }
    \over
    {
     \theta
     \left[
      \begin{array}{c}
       0 \\ \gamma/3
      \end{array}
     \right](-1/2|\tau)
    }
   }
  \right)^2
 \right\}.
\nonumber
\end{eqnarray}

These are the combination of generalized theta functions
 which appear in open string one-loop vacuum amplitude.

\end{document}